# Interaction-driven spontaneous broken-symmetry insulator and metals in ABCA tetralayer graphene

Kai Liu[1#], Jian Zheng[1#], Yating Sha[1], Bosai Lyu[1], Fengping Li[2], Youngju Park[2], Yulu Ren[1], Kenji Watanabe[3], Takashi Taniguchi[4], Jinfeng Jia[1], Weidong Luo[1], Zhiwen Shi[1], Jeil Jung[2,5*], Guorui Chen[1*]

[1]*Key Laboratory of Artificial Structures and Quantum Control (Ministry of Education), Shenyang National Laboratory for Materials Science, School of Physics and Astronomy, Shanghai Jiao Tong University, Shanghai 200240, China.*

[2]*Department of Physics, University of Seoul, Seoul 02504, Korea.*

[3]*Research Center for Electronic and Optical Materials, National Institute for Materials Science, 1-1 Namiki, Tsukuba 305-0044, Japan*

[4]*Research Center for Materials Nanoarchitectonics, National Institute for Materials Science, 1-1 Namiki, Tsukuba 305-0044, Japan*

[5]*Department of Smart Cities, University of Seoul, Seoul 02504, Korea.*

[#]These authors contributed equally to this work.

*Correspondence to: chenguorui@sjtu.edu.cn, jeiljung@uos.ac.kr

**Interactions among charge carriers in graphene can lead to the spontaneous breaking of multiple degeneracies. When increasing the number of graphene layers following rhombohedral stacking, the dominant role of Coulomb interactions becomes pronounced due to the significant reduction in kinetic energy. In this study, we employ phonon-polariton assisted near-field infrared imaging to determine the stacking orders of tetralayer graphene devices. Through quantum transport measurements, we observe a range of spontaneous broken-symmetry states and their transitions, which can be finely tuned by carrier density $n$ and electric displacement field $D$. Specifically, we observe a layer antiferromagnetic insulator at $n = D = 0$ with a gap of approximately 15 meV. Increasing $D$ allows for a continuous phase transition from a layer antiferromagnetic insulator to a layer polarized insulator. By simultaneously tuning $n$ and $D$, we observe isospin polarized metals, including spin-valley-polarized and spin-polarized metals. These transitions are associated with changes in Fermi surface topology and are consistent with the Stoner criteria. Our findings highlight the efficient fabrication of specially stacked multilayer graphene devices and demonstrate that crystalline multilayer graphene is an ideal platform for investigating a wide range of broken symmetries driven by Coulomb interactions.**

When Coulomb interaction dominates over kinetic energy, the degeneracies of spin and valleys in graphene can be lifted, and the competitions among energies of different degrees of freedom can lead to rich broken-symmetry phases[1–9]. For monolayer graphene, however, the linear energy dispersion hinders the spontaneous broken-symmetries due to the large kinetic energy (Fig. 1a). Although the application of a strong magnetic field or the twisting of graphene layers can reduce kinetic energy[10–15], studying the physics of the Landau levels or the moiré flatbands necessitates the consideration of additional complex energy factors or the involvement of large unit cells, respectively. These complexities go beyond the scope of the simple intrinsic graphene model. In this regard, an alternative and more straightforward approach to enhance Coulomb interactions is by increasing the number of layers through rhombohedral stacking[16–21].

Rhombohedral graphene, characterized by an ABCABC... stacking sequence, exhibits a dispersion relation $E_k \sim p^N$, where $E_k$ is the kinetic energy, $p$ is the momentum, and $N$ is the number of layers[22–24]. Consequently, the kinetic energy can be significantly reduced in thicker graphene. Figure 1a and b illustrate the unit cell and the calculated band structures at K point of rhombohedral graphene, ranging from monolayer to ABCA tetralayer graphene (ABCA-4LG). The presence of interlayer hopping results in pronounced trigonal warping and enhanced van Hove singularities (VHS) near the charge neutrality point (CNP) as $N$ increases, as demonstrated by the line cut and density of states (DOS) in Fig. 1c and d. Considering the strong Coulomb interactions, the substantial DOS at CNP and VHS can lead to instabilities of the Fermi circles, potentially giving rise to new ground states with broken symmetries. The band structure and correlations of rhombohedral graphene can be significantly tuned by a vertical electric displacement field $D$, which contributes to the energy difference $\Delta$ between top and bottom layers, as shown in Fig. 1f and g. Specifically, ABCA-4LG can open a gap at CNP along with modification in its correlation, layer polarization, and VHS.

Based on the calculated DOS derived from single-particle band structures, we initially anticipated that ABCA-4LG would exhibit lower resistivity (higher conductivity) compared to thinner layers due to its larger DOS. However, our experimental findings in Fig. 1e demonstrate a significantly higher resistivity for ABCA-4LG compared to monolayer (A), bilayer (AB), and trilayer (ABC) graphene at CNP ($V_g$ = 0), when the temperature $T$ = 1.5 K. It is important to note that all the graphene layers presented in Fig. 1e are on the hBN substrate, which mitigates qualitative differences in electron-hole puddles, charge mobility, and dielectric screening effects. Consequently, the observed giant resistivity peak in ABCA-4LG is deemed to be intrinsic in nature.

To systematically study the unexpected resistivity peak of ABCA-4LG, the high-quality dual gated device is necessary, as it enables independent tuning of $n$ and $D$. Earlier experiments of moiré-less ABCA-4LG systems had been carried out for suspended samples with limited accessible electron densities and electric fields[18] or with scanning probe microscopes for the larger ABCA domains in marginally twisted double bilayer graphene[25]. The fabrication of high-quality rhombohedral graphene devices thicker than bilayer poses a challenge. While rhombohedral stacking exists naturally as domains in bulk graphite and exfoliated thin layers, it is energetically metastable and easy to transition into Bernal stacking during the fabrication process such as the dry transfer. Therefore, it is crucial to effectively monitor the stacking orders throughout the fabrication process, particularly after hBN coverage, as traditional tools like Raman spectroscopy are not readily applicable in these cases.

To address this challenge, we employed a phonon-polariton assisted near-field optical imaging technique, which enabled the identification of rhombohedral and Bernal graphene across a coverage hBN flake. Such penetration imaging was conducted on a scanning near-field optical microscope (SNOM) system, known for its capability to image the stacking orders of on-surface graphene[26,27]. However, when graphene is covered by a hBN flake, conventional near-field optical imaging of graphene becomes inaccessible due to the highly localized near field and strong screening effect of the hBN flake. To image the embedded graphene, we carefully selected a specific excitation frequency that lies in one of the Reststrahlen bands of hBN (Fig. 2a). At this frequency, hBN behaves as a waveguide[28,29], transmitting the optical response of the graphene located beneath the hBN slab to its top surface, as illustrated in Fig. 2b (see *Methods* for details).

In our experimental approach, after the identification of exfoliated graphene layer numbers by optical contrast and SNOM, we used in-situ AFM cutting technique[30] to isolate the ABCA domains from ABAB stacks. Subsequently, these domains were encapsulated by hBN thin films. Following the encapsulation, both the ABCA and ABAB graphene domains become invisible in terms of topography and near-field optical imaging, as illustrated in Fig. 2c and d. To visualize the ABCA and ABAB domains, we finely adjusted the excitation frequency to match one of the Reststrahlen bands of hBN. This allowed the ABCA and ABAB domains to become distinguishable (Extended Data Fig. 1), with maximum contrast observed at 1550 cm$^{-1}$ (Fig. 2e). By phonon-polariton assisted near-field optical imaging technique, we successfully fabricated a few moiré-less hBN/ABCA-4LG/hBN devices with dual gates. A schematic of a typical device's side-view and an optical image are presented in Fig. 2f and g, respectively. The implementation of top and bottom gates provides independent control over $n$ and $D$ of the ABCA-4LG (see *Methods*).

Through quantum transport measurements, we have observed two distinct insulating phases at $D = 0$ and $D \neq 0$ at CNP ($n = 0$), as shown in Fig. 3a. The color plot represents the measured resistivity $\rho_{xx}$ as a function of $n$ and $D$ at $T = 1.5$ K. The insulating nature of these two phases is further confirmed by the temperature-dependent resistivity curves shown in Fig. 2b, which presents the line cut along $D$ at $n = 0$ for varied temperatures. From the Arrhenius plot of the temperature dependence of $\rho_{xx}$ (Extended Data Fig. 2), we can estimate a transport gap $\Delta$ of the insulator at $n = D = 0$ to be approximately 15 meV. Furthermore, we observe that this gap linearly deceases with $D$ and eventually vanishes at $|D| \approx 0.1$ V/nm, as shown in Fig. 3c. In the case of the insulator at $D \neq 0$, the gap emerges when $|D| > 0.15$ V/nm and linearly increases with $|D|$. These two insulating phases are connected by a low resistivity region spanning $|D| \sim 0.1$ to 0.15 V/nm. Notably, this low resistivity state exhibits metallic behavior at low temperatures when $T < 14$ K (Extended Data Fig. 2).

Similar to AB bilayer[16,31–33] and ABC trilayer graphene[17,34], the insulating state observed at large $|D|$ in ABCA-4LG can be attributed to a layer-polarized insulator (LPI) resulting from the broken inversion symmetry. The low-energy conduction and valence bands of ABCA-4LG arise from the orbitals of the two diagonal sublattices in the top and bottom layers. When a large $D$ is applied, as shown in the inset of Fig. 3c, the inversion symmetry of top and bottom sublattices is broken, the valence and conduction bands are contributed from different layers, leading to the opening of a LPI gap whose size is linearly dependent on $D$.

In contrast to AB bilayer and ABC trilayer cases, the insulating state observed at $D = 0$ in ABCA-4LG is unexpected based on the single-particle band structures, which predict the presence of finite DOS at zero energy. The dashed line in Fig. 3c represents the calculated single-particle gap at different $D$ values. Interestingly, the observed unexpected gap nearly emerges when the calculated gap is zero, indicating the existence of finite carriers arising from the overlap between valence and conduction bands in the single-particle picture. We argue that the insulating state observed at $n = D = 0$ originates from a layer antiferromagnet (LAF, shown as the middle inset of Fig. 3c) state of the finite carriers in ABCA-4LG.

Theoretical predictions have identified all possible insulating ground states with different broken symmetries in rhombohedral graphene. The layer pseudospin polarization, namely the layer resolved charge distribution of the four spin-valley flavors can be classified as flavor antiferromagnetic (AF) for the layer charge balanced configuration, the flavor ferrimagnetic (Fi) for partial, and flavor ferromagnetic (F) for full layer charge polarization[1,8]. These phases often have associated quantum Hall phases due to the Berry curvatures at the Dirac points and they include the quantum anomalous Hall (QAH), LAF, quantum spin Hall (QSH) for the layer charge balanced flavor AF phase, the quantum valley Hall (QVH) for the flavor F phase, and the ALL state for the flavor Fi phase[5,8]. However, we can firstly rule out QAH, QSH, and ALL states as they would exhibit quantized longitudinal and Hall resistivities, which are absent in our measurements. Additionally, QVH can be ruled out since it requires a charge-layer polarization, which corresponds to a non-zero $D$. Therefore, the only remaining plausible candidate is the LAF state with broken time reversal symmetry. We also performed mean-field Hartree Fock calculations, and find that the LAF phase is energetically favorable than Fi and F phase at $D = 0$, and phase transition from LAF to Fi, and to F phase occur when gradually increasing the electric fields (see *Methods* and Extended Data Fig. 4 for details).

The LAF insulating state is further supported by our magneto-transport measurements. We investigate the resistivity at $n = D = 0$ under an in-plane magnetic field $B_{//}$, and an out-of-plane magnetic field $B_\perp$. Figure 3d shows that the resistivity and the gap remain largely unchanged by $B_{//}$ (more data in Extended Data Fig. 3), while experiencing a significant drop with $B_\perp$. Since $B_{//}$ and $B_\perp$ has the same effect on the spin, the observed decrease in the gap under $B_\perp$ is attributed to the influence on the valley degree of freedom. To explain this behavior, we construct a gap diagram as shown in Fig. 3e, based on the LAF state. In the presence of $B_\perp$, the valley can be illustrated following the LAF state. Under a $B_\perp$, the valley Zeeman splitting leads to a reduction of the LAF gap by $g_v \mu B_\perp$, where $g_v$ represents the valley g-factor, and $\mu$ represents the Bohr magneton. Analyzing the measurement results in Figure 3d, we find that the reduction in the gap is approximately 11 meV at 12 T. This allows us to estimate the value of $g_v$ to be around 20, which is consistent with previous experimental findings in graphene[35].

The observation of the LAF insulator state has been reported in suspended bilayer[31,36], trilayer[17], and multilayer rhombohedral graphene[18,21]. However, in graphene on hBN, the bilayer and ABC trilayer do not exhibit such an insulating state, possibly due to the screening effect from the hBN dielectric environment. In the case of ABCA-4LG, the strong Coulomb interactions are sufficient to drive the emergence of the LAF state. The reproducibility of the LAF state in hBN-encapsulated ABCA-4LG is evident, as similar behaviors are observed in multiple devices (Extended Data

Fig. 5 and 6). It is worth noting that the magnetic field dependence of the LAF state in our study is opposite to that observed in suspended graphene. In the suspended samples, the LAF gap increases with $B_\perp$[17,18,31,36]. The contrasting magnetic field dependence observed in these two different sample structures raise intriguing questions for further investigation.

Doping in ABCA-4LG can lead to interaction-driven broken symmetries for itinerant carriers, potentially resulting in spontaneous spin and/or valley polarizations through a different mechanism, following the Stoner criterion, $D(E_F)U > 1$, where $D(E_F)$ is the DOS at the Fermi energy $E_F$, and $U$ is the Coulomb energy[37], compared to the insulating states at CNP. In ABCA-4LG, upon doping, $U$ is typically large and $D(E_F)$ at the VHS can be further tuned by $D$, providing a tunable platform of Stoner ferromagnetism. It was argued that the non-local exchange between electrons favors formation of electron or hole pockets of nearby electrons in $k$-space[38], leading in systems with finite $D$ and carrier densities to the so-called momentum-space exchange condensation that favors spontaneous spin-valley flavor polarization with an eventual onset of nematic broken rotational symmetry phases. In Fig. 4a, due to the presence of PN junctions formed by misalignment between the top and bottom gates at opposite signs, which leads to bad contact resistance, our focus is on the resistivities in the hole-side at negative $D$ and the electron-side at positive $D$. The metallic regions with distinct behaviors are separated by resistivity bumps, indicated by the white lines (representing resistivity humps) in Fig. 4a. In a perpendicular magnetic field $B = 3$ T, Landau levels with different degeneracies are developed, as shown in Fig. 4b.

To further investigate the degeneracies in different regions, in Fig. 4d, we performed direct measurements of quantum oscillations as a function of the perpendicular magnetic field $B$, while fixing the values of $n$ and $D$ in three different regions. Figure 4e shows the fast Fourier transformation (FFT) of the quantum oscillations in Fig. 4d, and clearly reveals one dominant peak for each region, located at $f = 1/4$, $1/2$ and $1$, respectively. These frequencies correspond to 4-fold, 2-fold, 1-fold degenerate bands. In graphene, the expected Landau level degeneracy is typically 4, considering the degeneracy of spin and valley. However, the observed degeneracies of 1 and 2 indicate a spin and valley polarized (SVP) state and a spin polarized (SP) state, respectively[38,39]. The anomalous Hall signal in the red curve of Fig. 4f serves as the signature of valley polarization in SVP metallic state. The spin polarization is supported by Fig. 4h, where the SP metallic state is stabilized, and its phase boundary shifts to higher densities under an in-plane magnetic field. The Landau levels between different polarized states are represented by slanted lines, indicating that the Landau levels depend on both *n* and *D* simultaneously. These regions correspond to the presence of electron and hole Fermi pockets, illustrated as II, IV and VI in Fig.4g.

Fig. 4c summarizes our experimental discoveries, presenting the broken-symmetry phase diagram of ABCA-4LG as a function of *n* and *D*:

1. At $n = D = 0$, the strong Coulomb interactions result in the breaking of time reversal symmetry and the formation of a LAF insulator.

2. At large *D*, the broken sublattice inversion symmetry leads to LPI.

3. Between LAF and LPI, there exists low-resistivity intermediate state (IS).

4. At $n \neq 0$ and $D \neq 0$, the Coulomb interactions break the spin and valley degeneracies, giving rise to the development of spontaneous Stoner ferromagnetic metal phases, namely SVP and SP metallic states. The Fermi surfaces corresponding to each of the broken-symmetry metallic phases are illustrated in Fig. 4g. The phase boundaries of SP and SVP regions can also be clearly identified in Landau level fan diagram in Extended Data Fig. 7.

The simple lattice structure of rhombohedral graphene provides a convenient platform for theoretical studies and offers better control over sample homogeneity and reproducibility compared to moiré superlattices. Our discovery of interaction-driven broken-symmetry insulators and metals in ABCA-4LG opens up new avenues for exploring correlations, magnetism, and topology in moiré-less 2D crystals. We have shown a pathway to systematically construct rhombohedral multilayer graphene devices that will have broad implications in the investigation of phenomena such as the quantum anomalous Hall effect resulting from a delicate balance between competing phases with different band topology and unconventional superconductivity with coexisting orbital and spin magnetism. The phonon-polariton assisted near-field optical imaging technique used in our study can also be applied to imaging other materials with hBN coverage, including twisted graphene, transition metal dichalcogenides, and other air-sensitive 2D crystals.

**Methods**

**Sample fabrications** Graphene, graphite and hBN are mechanically exfoliated on $SiO_2$(285nm)/Si substrates, and the layer numbers are identified using optical contrast and atomic force microscopy. The stacking order of trilayer and tetralayer graphene is identified using SNOM and Raman spectroscopy. A dry transfer method using polypropylene carbonate or polycarbonate is implemented to construct the heterostructures. Standard e-beam lithography, reactive ion etching and metal evaporation are conducted to make the devices into Hall bar geometry with the one-dimensional edge contacts. After each step of transfer and fabrications, SNOM imaging is performed to check the stacking orders of graphene.

**Phonon-polariton assisted near-field optical imaging** The key factor that enables the penetration imaging of the embedded graphene across hBN flake is the hyperbolic nature of hBN crystal. In the Reststrahlen band between $1,370-1,610$ cm$^{-1}$, the real part of the in-plane permittivity ($\epsilon_t$) of hBN is negative while that of the out of plane permittivity ($\epsilon_z$) is positive. As a result, hBN supports hyperbolic phonon polaritons that mainly propagate along specific direction at a fixed angle $\theta$ with respect to the z axis, as $\theta = \arctan(i\sqrt{\epsilon_t}/\sqrt{\epsilon_z})$, determined by the hyperbolic dielectric function. The advantage of the directional polaritonic rays is the greatly enhanced propagation length, due to reduced radical spreading loss. Previous study has reported subdiffractional focusing and guiding of the polaritonic rays in hBN[29]. Here in our experiment, the directionally propagating polaritonic rays carry the underneath graphene domain patterns to the hBN top surface, which finally coupled out by the metallic AFM tip and recorded by an MCT detector placed in the far field. As ABAB graphene domain typically possess larger optical conductivity than the ABCA graphene, therefore we can identify them from their near-field optical response by scanning them even with a hBN coverage. What's more, SNOM imaging enables high resolution down to 10 nm, which can much enhance the precision in device fabrication. As a result, the phonon-polariton assisted imaging method can efficiently help identify the graphene stacking order after transfer process and push the rhombohedral graphene related devices to a more applicable level.

**Transport measurements** The transport measurement is done in the 1.5 K base temperature Oxford variable temperature insert (VTI) system. Stanford research system SR830, SR860, and Guangzhou Sine Scientific Instrument OE1201 lock-in amplifiers with an alternating-current of 10 nA at a frequency of 17 Hz in combination with a 100-MΩ resistor are used to measure the resistivity. Keithley 2400 source meter is used to apply the gate voltages. The displacement field $D$ is set by $D = (D_b + D_t)/2$, and carrier density is determined by $n = (D_b - D_t)/e$. Here, $D_b = +\varepsilon_b (V_b - V_b^0)/d_b$, $D_t = -\varepsilon_t (V_t - V_t^0)/d_t$, where $\varepsilon$ and $d$ are the dielectric constant and thickness of the dielectric layers, respectively, $V_b^0$ and $V_t^0$ are effective offset voltages caused by environment-induced carrier doping.

**Band Calculation** The electronic structure of ABCA stacked multilayer graphene is modeled with a tight-binding Hamiltonian whose hopping terms are fitted to reproduce density functional theory energy bands obtained within the local density approximation. This Hamiltonian reads

$$H_{4LG} = \begin{pmatrix} H_{11} & H_{12} & H_{13} & H_{14} \\ H_{12}^\dagger & H_{22} & H_{23} & H_{24} \\ H_{13}^\dagger & H_{12}^\dagger & H_{33} & H_{34} \\ H_{14}^\dagger & H_{24}^\dagger & H_{34}^\dagger & H_{44} \end{pmatrix}$$

where $(H_{ll})_{2\times 2}$ represent the intralayer graphene Hamiltonian diagonal blocks for layers $l = 1, 2, 3, 4$, and the interlayer coupling $(H_{ij})_{2\times 2}$, $(i \neq j = 1, 2, 3)$ terms near the Dirac points are respectively given by

$$H_{ll}(\boldsymbol{k}) = \begin{pmatrix} u_{A_l} & v_o \pi^\dagger \\ v_o \pi & u_{B_l} \end{pmatrix} + V_{ll}\mathbb{I}, \; H_{12}(\boldsymbol{k}) = \begin{pmatrix} -v_4 \pi^\dagger & -v_3 \pi \\ t_1 & -v_4 \pi^\dagger \end{pmatrix}, \; H_{13}(\boldsymbol{k}) = \begin{pmatrix} 0 & t_2 \\ 0 & 0 \end{pmatrix}.$$

Here $\pi = (\nu p_x + i p_y)$ depends on the valley index $\nu = \pm 1$ where the momentum is measured from the Dirac points $K_\nu = (\nu \frac{4\pi}{3a}, 0)$, and for each valley. The positive valued parameters associated with the hopping terms are $v_i = \sqrt{3}a|t_i|/2\hbar$ and the term $V_{ll}\mathbb{I}$ is used to define the potential of each layer. We model the drop of interlayer potential $\Delta$ between consecutive layers due to a perpendicular electric field through $V = \Delta \left(\frac{3}{2}, \frac{1}{2}, -\frac{1}{2}, -\frac{3}{2}\right)$. The hopping parameters used in this work $(t_0, t_1, t_2, t_3, t_4)$ = $(-2.6, 0.3561, -0.0093, 0.293, 0.144)$ eV are similar to those in Ref. [40] that are respectively the nearest neighbor intralayer hopping term between $A_l \& B_l$ sites, and interlayer $B_l \& A_{l+1}$, $A_l \& B_{l+2}$, $A_l \& B_{l+1}$, $A_l(B_l) \& A_{l+1}(B_{l+1})$ sites (Extended Data Figure 8). The diagonal site potentials $u_{A_l}(u_{B_l})$ at each sublattice of 4LG are $u_{A1} = u_{B4} = 0$, $u_{B1} = u_{A4} = 0.0122$ eV, $u_{A2/A3} = u_{B2/B3} = -0.0164$ eV.

**Spontaneous degeneracy breaking** We carry out a self-consistent Hartree-Fock (HF) calculation in the sublattice-spin basis $\lambda = (\kappa, \sigma)$ where $\kappa = \{A_1, B_1, A_2, B_2, A_3, B_3, A_4, B_4\}$ for the sublattice site labels and $\sigma = \{\uparrow, \downarrow\}$ for up/down spins. The HF Coulomb term $V_{\text{HF}}$ to be added in the band Hamiltonian $H_{4LG}$ can be written as

$$V_{\text{HF}} = \sum_{k,\lambda,\lambda'} \sum_{k'} [<k\lambda\, k'\lambda'|V|k\lambda\, k'\lambda'><c^\dagger_{k'\lambda}c_{k'\lambda}>c^\dagger_{k\lambda}c_{k\lambda}$$
$$- <k\lambda\, k'\lambda'|V|k'\lambda\, k\lambda'><c^\dagger_{k'\lambda'}c_{k'\lambda}>c^\dagger_{k\lambda}c_{k\lambda'}],$$

where the details of the calculations follow closely previous works[8,34]. Here, we have chosen $k$-point densities near the valleys equivalent to $576 \times 576$ and relative permittivity $\varepsilon_r = 4$.

We define energy terms based on the unrestricted HF equations[41], where the total energy ($E_{\text{tot}}$) consists of the tight-binding ($E_{\text{band}}$), Hartree ($E_{\text{H}}$), and exchange ($E_{\text{X}}$) energy terms,

$$E_{\text{tot}} = E_{\text{band}} + E_{\text{H}} + E_{\text{X}},$$

$E_{\text{H}} = \tfrac{1}{2}\sum_{\kappa\kappa'}\sum_{k\sigma}\sum_{k'\sigma'} <k(\kappa,\sigma)\,k'(\kappa',\sigma')|V|k(\kappa,\sigma)\,k'(\kappa',\sigma')><c^\dagger_{k'(\kappa',\sigma')}c_{k'(\kappa',\sigma')}><c^\dagger_{k(\kappa,\sigma)}c_{k(\kappa,\sigma)}>$,

$E_{\text{X}} =$

$-\tfrac{1}{2}\sum_{\kappa\kappa'}\sum_{k\sigma}\sum_{k'\sigma'}<k(\kappa,\sigma)k'(\kappa',\sigma')|V|k'(\kappa,\sigma)k(\kappa',\sigma')><c^\dagger_{k'(\kappa',\sigma')}c_{k'(\kappa,\sigma)}><c^\dagger_{k(\kappa,\sigma)}c_{k(\kappa',\sigma')}>\delta_{\sigma\sigma'}$,

$E_{\text{band}} = \left(\sum_{nk\sigma}\epsilon_{nk\sigma}<c^\dagger_{nk\sigma}c_{nk\sigma}>\right) - 2(E_{\text{H}} + E_{\text{X}}).$

The first term in $E_{\text{band}}$ is the orbital energy defined by the eigen energies $\epsilon_{nk\sigma}$ and eigen states $|nk\sigma>$ over the band index $n$, $k$ points, and spin $\sigma$.

We could classify the different ground states as the spin-valley flavor antiferromagnetic (AF) phase with equal charge polarization at opposite surface layers, the flavor ferrimagnetic (Fi) with partially unbalanced layer charge polarizations, and flavor ferromagnetic (F) states with maximal charge polarization difference between the top and bottom layers[1,8].

In Extended Data Fig. 4a, we show schematic layer polarization for each phase, where two of the four spin-valley flavors polarize the states to the top layer in AF phases which includes QAH and LAF, three for Fi, and four for F phases. We note that the Fi phase is an intriguing one with associated spin, charge, and valley Hall conductivities simultaneously.

Within HF theory for charge neutral bilayer[8] and trilayer[34], it was shown that AF phases, especially layer antiferromagnetic (LAF) phase, are energetically preferred over Fi and F phases at zero electric field. Application of a perpendicular electric field leads to phase transitions from the charge balanced AF to the unbalanced Fi, and F phases.

To examine the ground states for the charge neutral ABCA-4LG system, we compare the total energies for each phase, where we have used a 576×576 k point mesh grid near two Dirac points, $K$ and $K'$, and 18×18 k point grid for the rest of the first Brillouin zone, as seen in Extended Data Fig. 4b.

We verified that the charge neutral ABCA-4LG system also shows that the LAF phase are energetically favorable than Fi and F phase in the absence of external electric field, and the phase transition from LAF to Fi, and to F phase occur when gradually increasing the electric fields.

In Extended Data Fig. 4c-f, we illustrate the electric field dependent variations for all four phases on the band gap, total energy, band and Hartree energy, and exchange energy.

**Author contributions**
G.C. supervised the project. K.L. and J.Z. fabricated devices and performed transport measurements with the assistance of Y.S. and B.L. K.L. and Z.S. performed near-field infrared and AFM measurements. K.L., F.L., Y.R., Y.P., W.L. and J.J. calculated the band structures. K.W. and T.T. grew hBN single crystals. K.L., J.Z., Y.S., J.J. and G.C. analyzed the data. K.L., Z.S., J.J. and G.C. wrote the paper with input from all authors.

**Acknowledgment**
We acknowledge helpful discussions with Jianpeng Liu, Shiyong Wang, Yuanbo Zhang, and Feng Wang. This work is supported by National Key Research Program of China (grant nos. 2020YFA0309000, 2021YFA1400100, 2021YFA1202902, 2022YFA1402401), NSF of China (grant nos.12174248, 12074244) and SJTU NO. 21X010200846. G.C. acknowledges the sponsorship from Yangyang Development

Fund. K.W. and T.T. acknowledge support from the JSPS KAKENHI (Grant Numbers 20H00354, 21H05233 and 23H02052) and World Premier International Research Center Initiative (WPI), MEXT, Japan.

**Author Information**

The authors declare no competing financial interests. Correspondence and requests for materials should be addressed to G.C. (chenguorui@sjtu.edu.cn) and J.J. (jeiljung@uos.ac.kr).

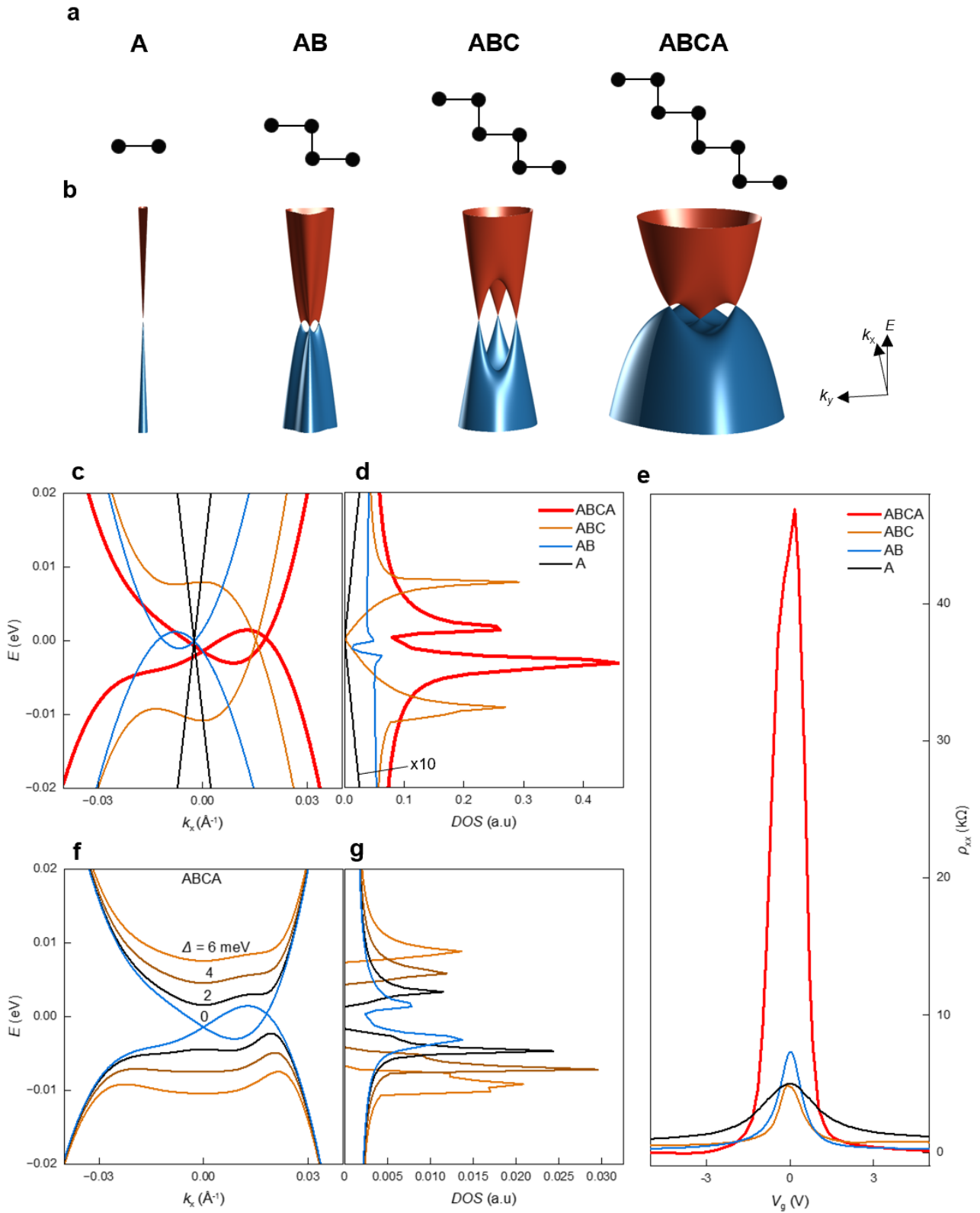

**Figure 1. Rhombohedral graphene family. a** and **b,** Unit cells and band structures of monolayer, AB bilayer, ABC trilayer and ABCA tetralayer graphene. Conduction and valence bands are shown in red and blue, respectively. **c** and **d,** Line-cuts of band structures along $k_x$ direction when $k_y = 0$ and density of state (DOS) versus energy from monolayer to tetralayer. The value of DOS of monolayer graphene is amplified by a factor of 10 in **d** for visibility. **e,** Experimentally measured gate-dependent resistivity of different graphene layers in similar sample qualities. Data were obtained at $T = 1.5$ K. **f** and **g,** Band structures and DOS of ABCA-4LG with applied interlayer energy differences.

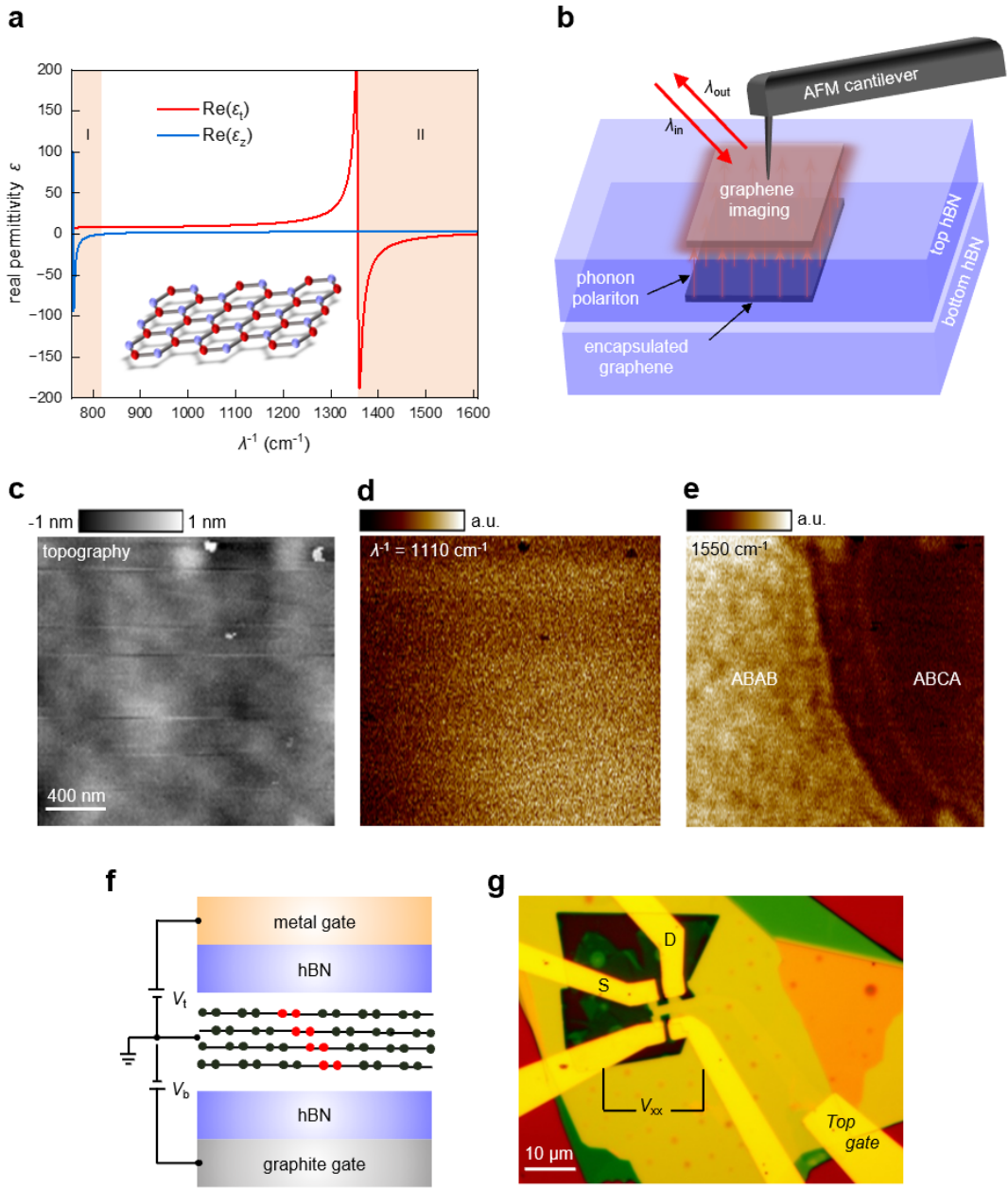

**Figure 2. Phonon-polariton assisted near-field optical imaging of graphene stacking orders under hBN coverage. a,** Derived real part of the permittivity tensor components of hBN. Type I lower and II upper Reststrahlen bands are shaded. Inset: a schematic of hBN crystal structure. **b,** Illustration of phonon-polariton assisted near-field optical imaging measurement of graphene encapsulated by hBN. Infrared light (red arrows) at certain wavenumber within the hBN Restsranhlen band is focused onto the apex of a metal-coated AFM tip. The phonon-polaritons (orange arrows) propagate onto top hBN surface, are scattered by the AFM tip and detected by an HgCdTe detector in the far field. **c-e,** Topography, SNOM images when incident wavenumbers $\lambda^{-1}$ = 1110 cm$^{-1}$ and 1550 cm$^{-1}$ of a hBN encapsulated tetralayer graphene, respectively. **f,** Schematic side view of the dual-gated hBN/ABCA-4LG/hBN Hall bar device. One unit cell of ABCA-4LG is labeled by red. **g,** Optical image of the final dual-gated hBN/ABCA-4LG/hBN Hall bar device. For transport measurement, we applied a current $I$ through source (S) and drain (D), and measure the voltage $V_{xx}$.

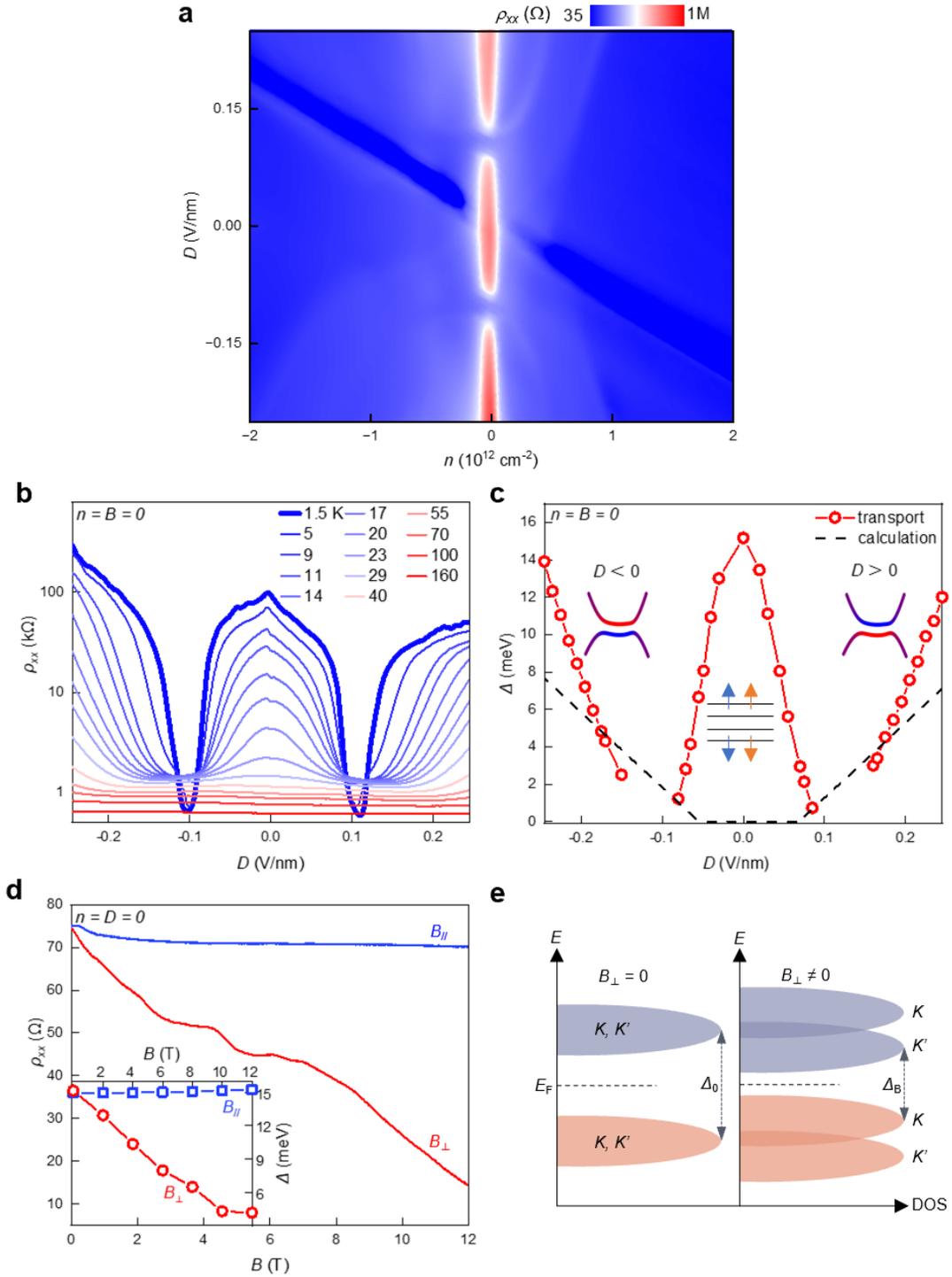

**Figure 3. Broken-symmetry insulators at $n = 0$. a,** Color plot of resistivity $\rho_{xx}$ as a function of carrier density $n$ and displacement field $D$. The color bar is in the log scale. **b,** $D$-dependent $\rho_{xx}$ for $n = 0$ at varied temperatures from 1.5 K to 160 K. **c,** $D$-dependent transport gap (red circles) extracted from Arrhenius plot of **b** and calculated single-particle gap (dashed line). The calculated gap size is reduced by a factor of 5.6 considering the screening in graphene. Inset at $D = 0$ illustrates the isospin flavors distribution of LAF state, four black lines represent ABCA-4LG, arrows represent spins, and two colors represent two valleys. Insets on two sides are band structures color-coded with red and blue, respectively, denoting wave functions in the top and bottom two layers. **d** Resistivity as a function of in-plane and out-of-plane magnetic fields. Inset shows the extracted transport gaps at different magnetic fields. **e,** Illustration of the LAF insulator and its valley Zeeman splitting in the out of plane magnetic field.

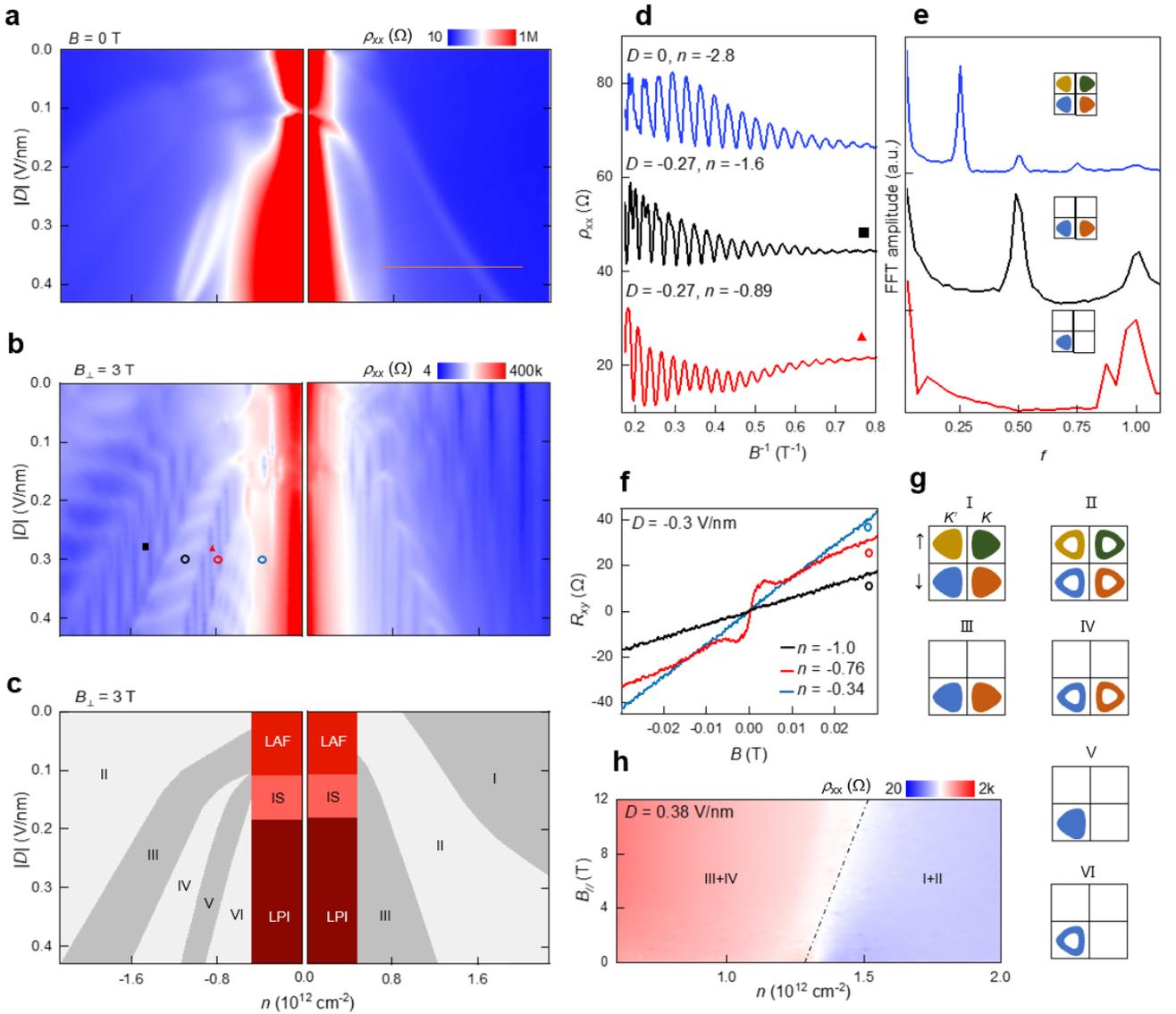

**Figure 4. Broken-symmetry metals in ABCA-4LG. a** and **b**, $n$ - $D$ color plot of resistivity at $B = 0$ T and $B = 3$ T, respectively. The color bars are linear in **a** and logarithmic in **b**. **c**, Experimental phase diagram of the broken-symmetry states. IS corresponds to intermediate state. **d**, SdH oscillations at normal, SP and SPV metal states. The data of SP and SVP metals are taken at the square (black) and triangle (red) in **b**. **e**, FFT analysis of the quantum oscillations in **d**. Inset: the corresponded Fermi surface contours of the three metals. **f,** Hall resistivity $R_{xy}$ as a function of magnetic field measured at black, red and blue circles in **b**. All the data were obtained at $T = 1.5$ K. **g,** Schematic Fermi surface contours of each phase defined in **c**. Filled colors represent the spin and valley flavors, and the inset of **e** follows the same convention. **h,** $n$ - $B_{//}$ color plot of resistivity at $D = 0.38$ V/nm. The data is taken along the horizontal orange line in **a**.

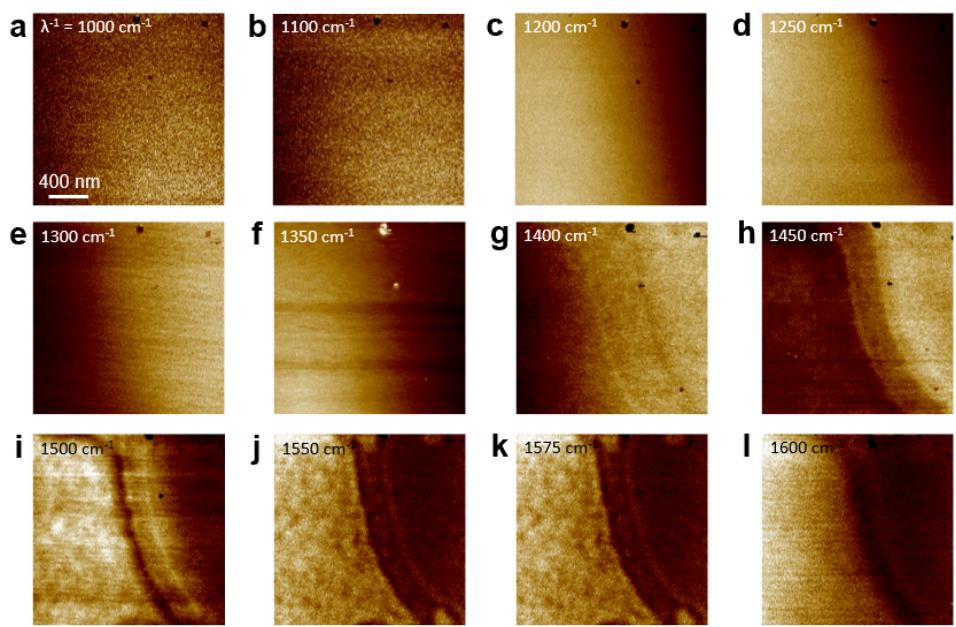

**Extended Data Figure 1 | Wavelength dependent SNOM contrast of ABCA and ABAB regions. a ~ l,** IR-SNOM images of the same tetra-layer graphene sample with that in Fig. 2c~e at different incident wavelength ranging from 1000 to 1600 cm$^{-1}$. Only when the incident wavelength lies in the hBN Reststrahlen (1370~1610 cm$^{-1}$), ABAB and ABCA regions can be clearly distinguished. Between 1500 ~ 1600 cm$^{-1}$, ABCA is always darker than ABAB (i ~ l), between 1400 ~ 1500 cm$^{-1}$, ABAB should be darker (Fig. g, h). Below 1350 there is no obvious boundary between ABCA and ABAB and the contrast is week. The hBN thickness is about 35 nm, and the laser source is a commercial quantum cascade laser (QCL) with output wavelength ranging from 900 cm$^{-1}$ to 1680 cm$^{-1}$.

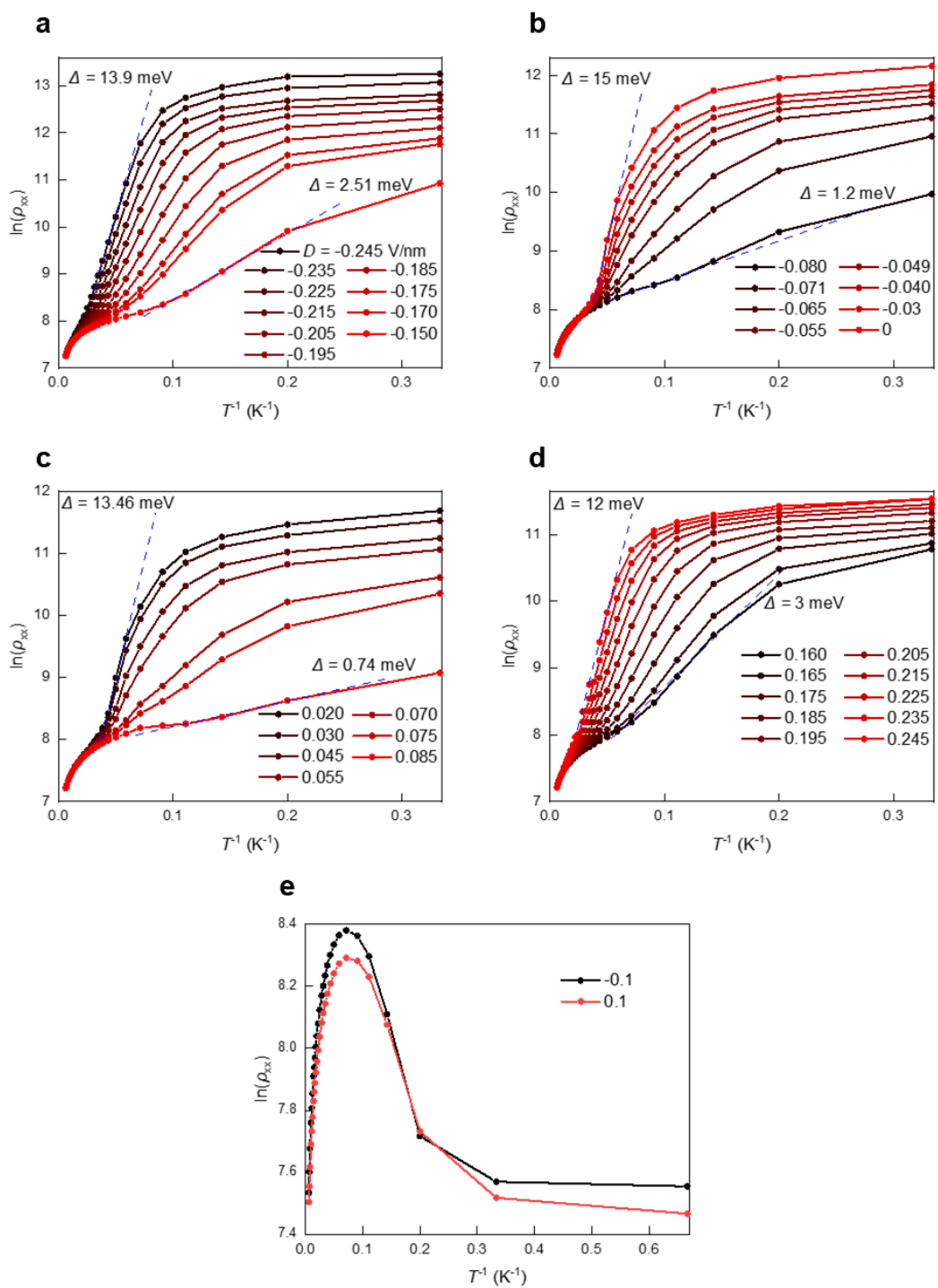

**Extended Data Figure 2 | Arrhenius plot.** $\ln(\rho_{xx})$ versus temperature at different displacement fields. The transport gap $\Delta$ in Fig. 3c is extracted according to thermal activation equation $\rho_{xx} \propto e^{-\Delta/2k_BT}$. The dashed lines in **a** to **d** represent the linear regions to estimate the gap. **a** and **d** correspond to LPI, **b** and **c** correspond to LAF, **e** correspond to IS which shows the metallic behavior when temperature is lower than 14 K.

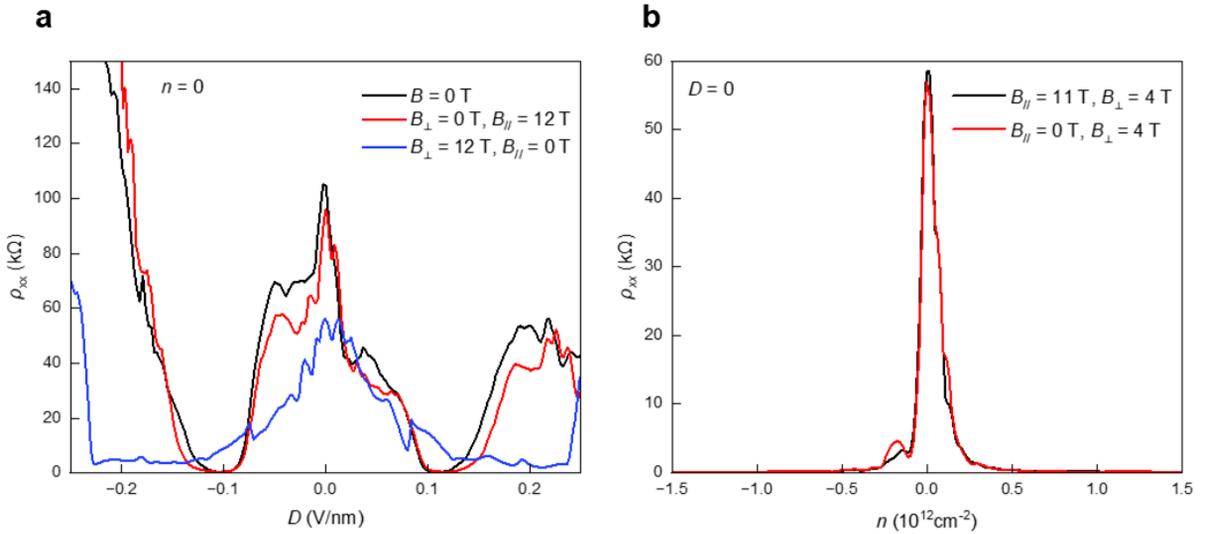

**Extended Data Figure 3 | Comparison of the response to the vertical and parallel magnetic field focusing around $D = 0$. a,** $\rho_{xx}$ - $D$ plot at $B = 0$ T; $B_\perp = 0$ T, $B_{//} = 12$ T and $B_\perp = 12$ T, $B_{//} = 0$ T. **b, b,** $\rho_{xx}$ - $n$ plot at $B_\perp = 4$ T, $B_{//} = 11$ T and $B_\perp = 4$ T, $B_{//} = 0$ T. Both data show parallel magnetic field have little influence on the correlated insulator state around CNP. A rotating probe is used to relatively change the direction of magnetic field. According to the measurement of hall resistance, pure in-plane or out of plane magnetic field (error within 0.1°) can be applied to sample.

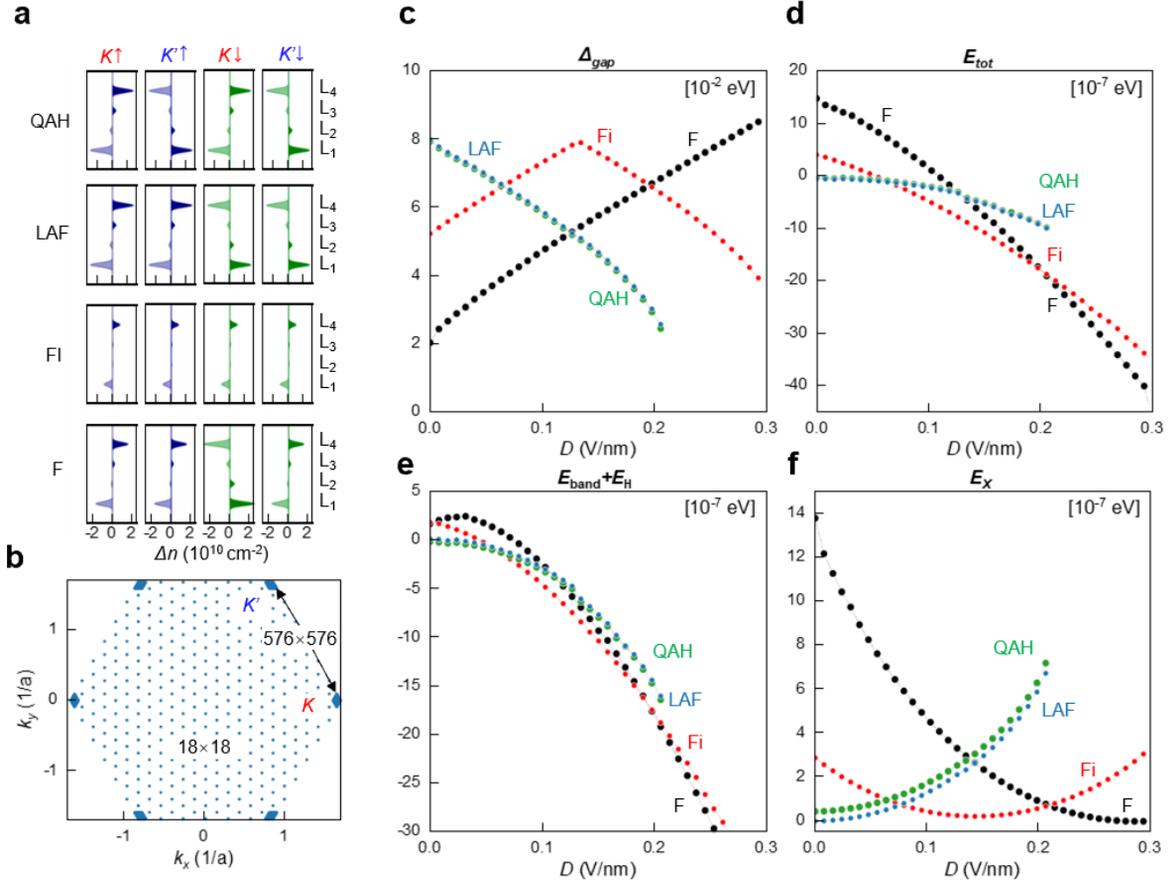

**Extended Data Figure 4 | Spontaneous degeneracy breaking. a** Layer resolved distribution of electrons and holes for charge neutral QAH, LAF, and flavor Fi, and F phases illustrating the layer polarization of the four spin-valley flavors ($K\uparrow$, $K'\uparrow$, $K\downarrow$, $K'\downarrow$). **b** k-point mesh grid of the first Brillouin zone where the dense grid near two Dirac points, $K$ and $K'$, is equivalent to 576×576 k points, and the coarse grid for the rest of the area is to 18×18 $k$ points. We show electric field dependent variations on **c** band gap ($\Delta_{gap}$), **d** total energy ($E_{tot}$), **e** the sum of band ($E_{band}$) and Hartree ($E_H$) energy, and **f** exchange energy ($E_X$) for each phase.

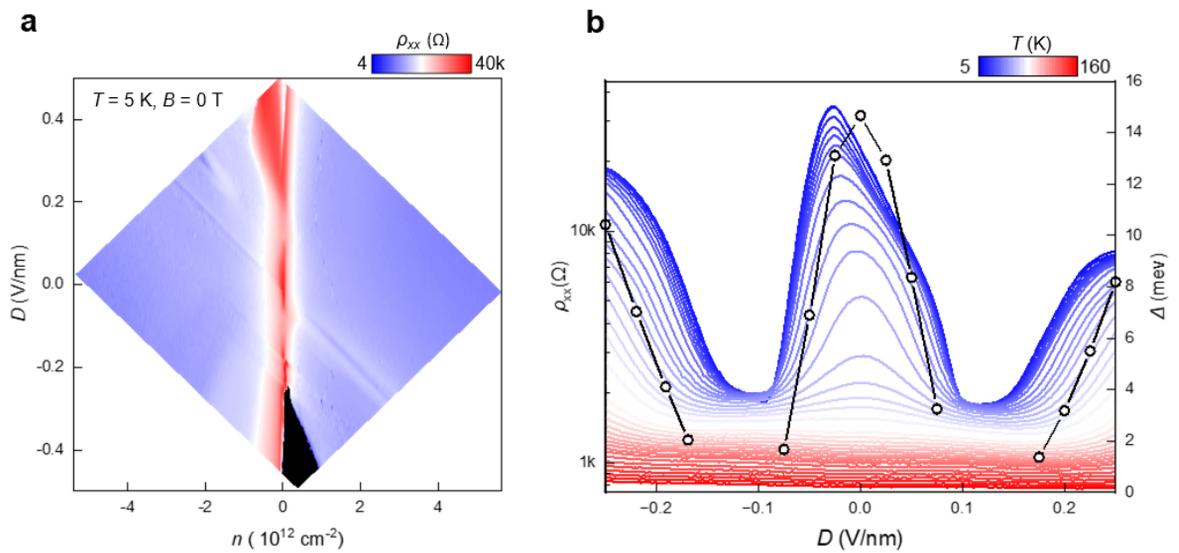

**Extended Data Figure 5 | Correlated insulating state at device 2. a,** $\rho$ - $n$ - $D$ color plot of device 2, it shows a log color scale from 4 to 40 k$\Omega$. **b,** Left axis corresponds to plot of $\rho_{xx}$ at different $D$ and fixed $n$ and $B$ of 0 for temperatures ranging from 1.5 K to 160 K. Right axis is the corresponding displacement field dependence of gap $\Delta$. Device 2 gives almost the same gap with device 1 in the main text.

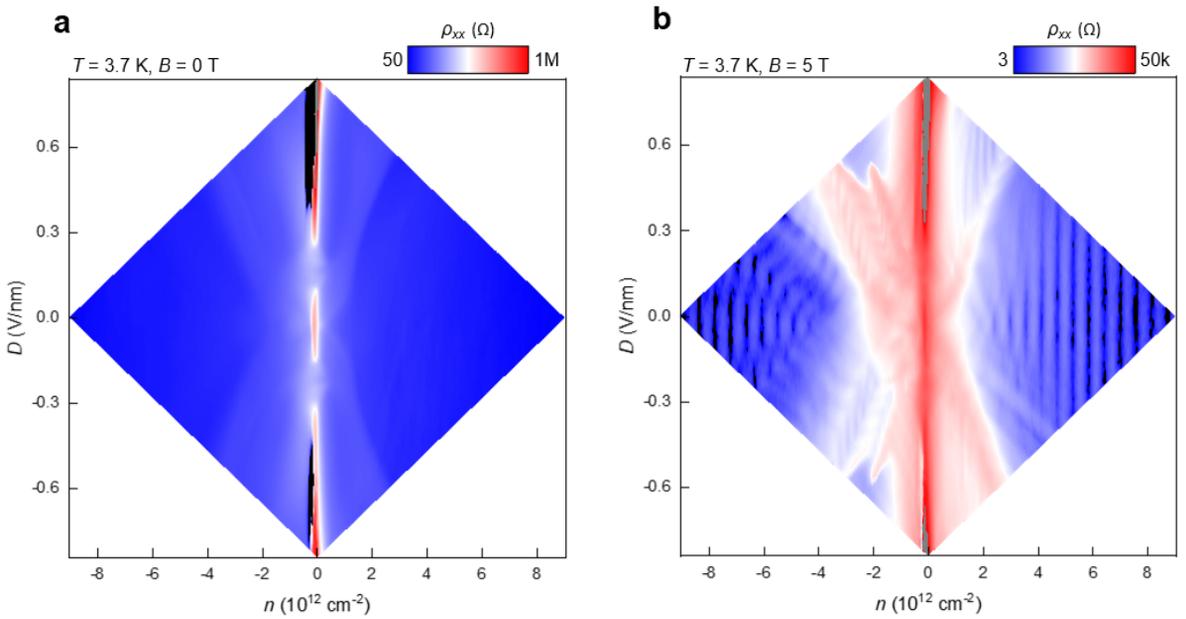

**Extended Data Figure 6 | Correlated insulating state and SP & SVP states at device 3. a,** $\rho$ - $n$ - $D$ color plot of device 3 when temperature $T$ = 3.7 K, it shows a log color scale from 50 to 1 MΩ. **b,** Corresponding $R$ - $n$ - $D$ color plot when $B$ = 5 T. SP and SVP occur at the similar regions with device 1 in the main text.

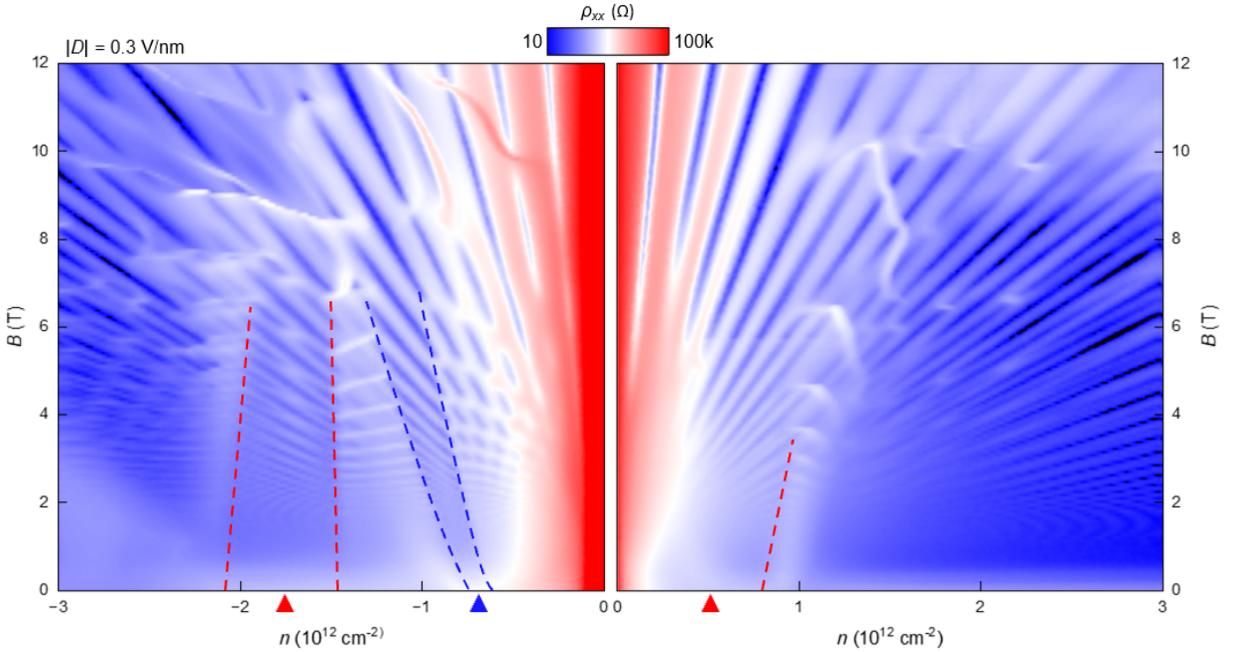

**Extended Data Figure 7 | Landau level fan of device 1 (the same sample in the main text).** $\rho - n - B$ color plot of device 1 at $D = 0.3$ V/nm when $T = 1.5$ K. SP and SVP states are also prominent in the landau level fan diagram which is indicated by blue and red triangles, and their area would develop with magnetic field since the Zeeman energy of spin and orbital should be changed (red and blue dashed lines show the boundary of SP and SVP. above the dashed line, all the degeneracy is open by the large magnetic field). The area of the SP metal indicted by the left hand side red triangle is almost unchanged, we ascribe it as a fully spin polarized region but the SP metal indicted by the right hand side red triangle as a partially spin polarized region.

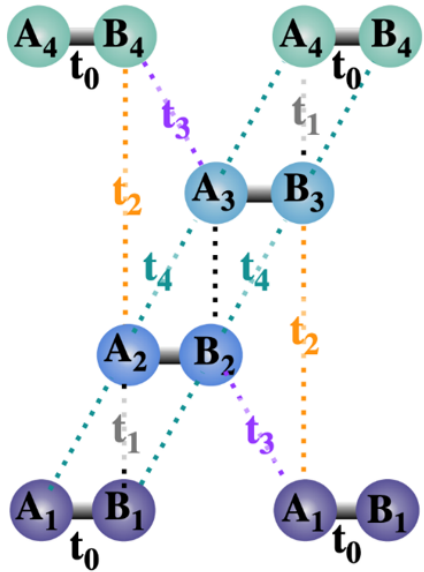

**Extended Data Figure 8 | Illustrations of hopping parameters.** Illustration of the intralayer and interlayer tight-binding hopping terms of ABCA multilayer graphene used in this work.